\documentclass{uemura}
\begin {document}
\title{\bf What can we learn from comparison between cuprates and
He films ? : phase separation and fluctuating superfluidity\/}  
\author{Y.~J.~Uemura}
\address{Dept. of Physics, 
Columbia University, New York, NY 10027, U.~S.~A.
\\E-mail: tomo@kirby.phys.columbia.edu}
\maketitle
\abstracts{
In the underdoped, overdoped, Zn-doped or stripe-forming regions of 
high-$T_{c}$ cuprate superconductors (HTSC), the superfluid density 
$n_{s}/m^{*}$ at $T\rightarrow 0$ shows universal correlations with $T_{c}$.
Similar strong correlations exist between 2-dimensional 
superfluid density and superfluid transition temperature 
in thin films of $^{4}$He in non-porous or porous media, and $^{4}$He/$^{3}$He
film adsorbed on porous media.
Based on analogy between HTSC and He film systems,
we propose a model for cuprates where: (1) the overdoped region is characterized
by a phase separation similar to $^{4}$He/$^{3}$He; and
(2) pair (boson) formation and fluctuating superconductivity occur at
separate temperatures above $T_{c}$ in the underdoped region.}

The magnetic field penetration depth $\lambda$ of superconductors
is related to the superconducting
carrier density $n_{s}$ divided by the effective mass $m^{*}$, as
$1/\lambda^{2} \propto n_{s}/m^{*}$.  In this paper, we shall 
refer to $n_{s}/m^{*}$ as the ``superfluid density''.
Since the discovery of HTSC, we have performed 
muon spin relaxation ($\mu$SR) studies of under- to optimally doped [1,2],
overdoped [3], and Zn-doped [4] cuprates, as well as HTSC systems associated
with the 
formation of static spin stripes [5,6].  In all of these systems,
we found strong correlations between $n_{s}/m^{*}$ at $T\rightarrow 0$ and
$T_{c}$, as shown in Fig. 1.  This figure suggests that the superfluid 
density is likely a crucial determining factor for $T_{c}$ of all these HTSC
systems.  These correlations in the 
underdoped region have been interpreted
in terms of Bose-Einstein (BE) to BCS crossover [7-9], 
phase fluctuations
[10], XY-model [11], as well as via RVB-type [12] pictures.

\begin{figure}[t]
\begin{center}
\epsfxsize=22pc 
\rotatebox{90}{\epsfbox{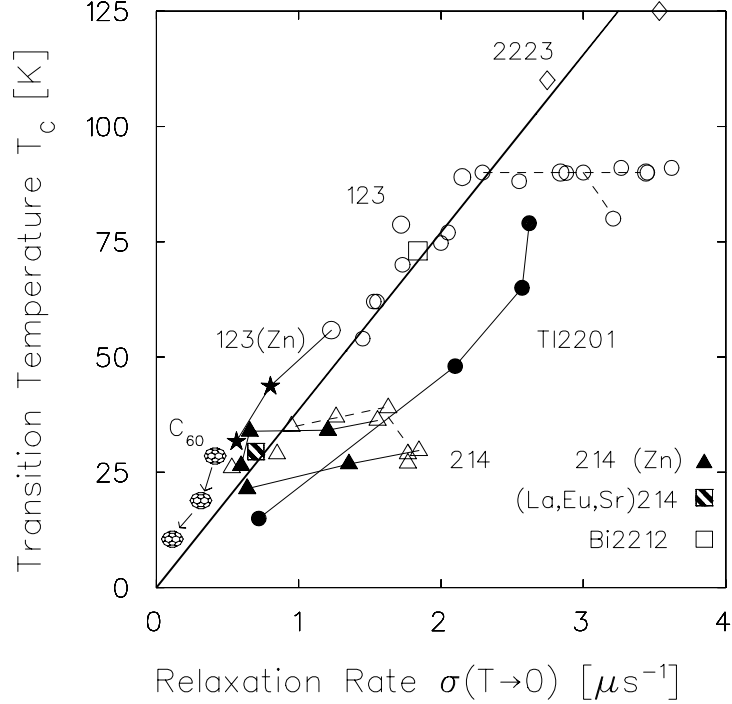}}
\caption
{Superconducting transition temperature 
$T_{c}$ of HTSC systems plotted versus
muon spin relaxation rate $\sigma(T\rightarrow 0)\propto 1/\lambda^{2}
\propto n_{s}/m^{*}$ [1-5]. Y123 systems on the solid line
are in the underdoped region, while Tl2201 systems are in the  
overdoped region.}
\end{center}
\end{figure}

Thin films of $^{4}$He also exhibit strong correlations between 
their two-dimensional superfluid area density $n_{s2d}/m^{*}$ and
$T_{c}$.  In Figure 2, we replot published results of 
$^{4}$He film on mylar sheet (non-porous media) [13],  
vycor glass (porous media) [14,15], as well as 
thin film $^{4}$He/$^{3}$He mixture on alumina powder 
(porous media) [16].
The horizontal axis was obtained 
after converting the He coverage into 2-dimensional (2-d) areal boson density 
divided by the boson mass $n_{b2d}/m_{b}$, and then into 
the 2-d Fermi temperature $T_{F2d}$ assuming $n_{b2d} = n_{s2d}/2$
and $m_{b} = 2m^{*}$.  The Kosterlitz-Thouless (KT) transition temperature
$T_{KT} = T_{F2d}/8$ for the strong coupling limit [17] is shown by the solid
line.  $T_{c}$ scales with $T_{KT}$, as expected for 
an ideal Bose gas composed of tightly-bound fermion pairs 
in a pure 2-d environment.

\begin{figure}[t]
\begin{center}
\epsfxsize=20pc 
\rotatebox{90}{\epsfbox{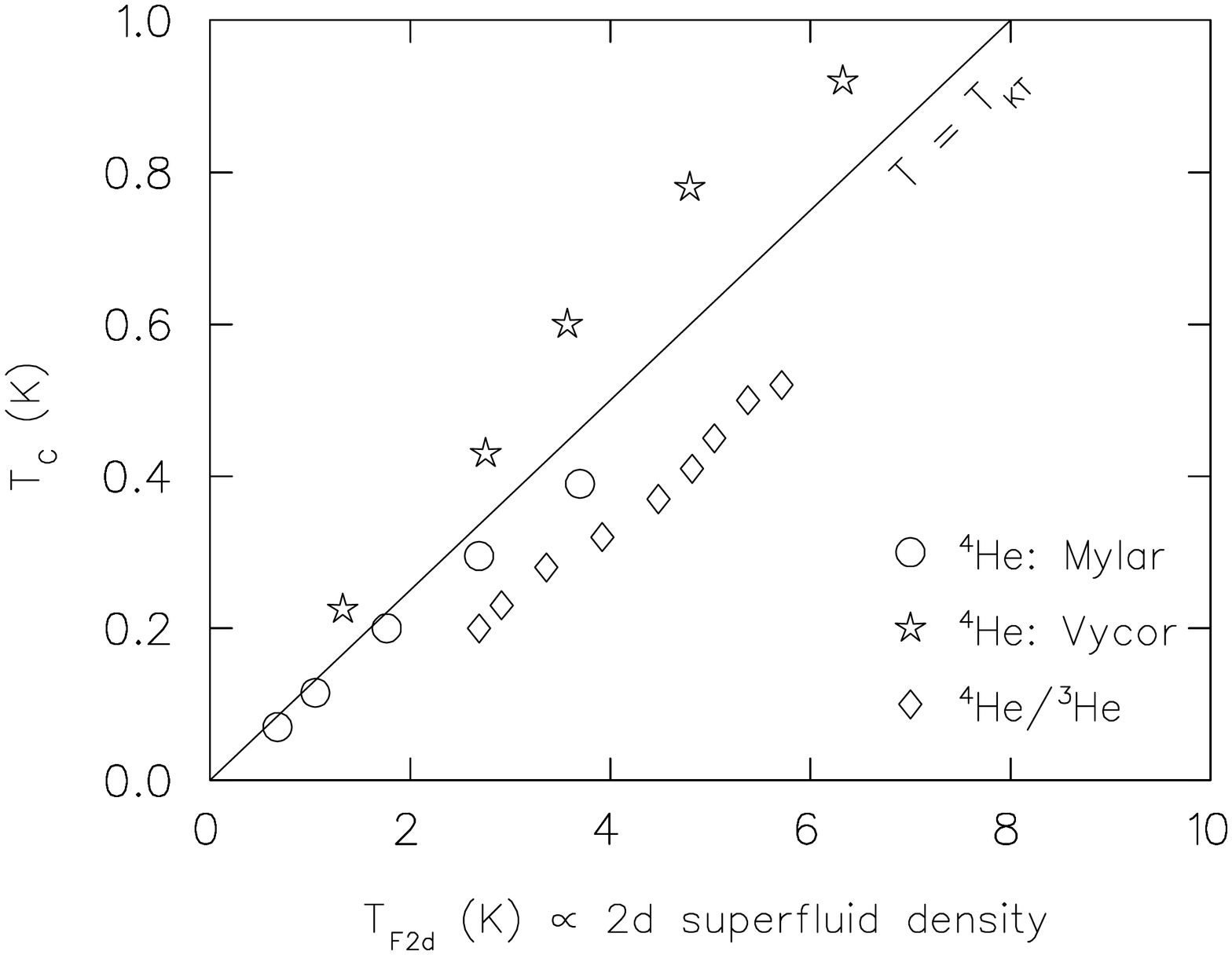}} 
\caption
{Superfluid transition temperature $T_{c}$ of $^{4}$He
film adsorbed on Mylar film [13], porous Vicor glass [14,15] and
$^{4}$He/$^{3}$He mixture adsorbed on fine alumina powder [16],
plotted versus 2-d superfluid density at $T\rightarrow 0$.  The horizontal
axis is shown by converting 2-d boson density $n_{b2d}$ and mass $m_{b}$ into
fermionic language as $n_{b2d}=n_{s2d}/2$ and $m_{b}=2m^{*}$ and then
calculating the corresponding 2-d 
Fermi temperature $T_{F2d} \propto n_{s2d}/m^{*}$.  The solid line
indicates the superfluid density expected at the Kosterlitz-Thouless transition
temperature $T_{KT}$.}
\end{center}
\end{figure}

Figures 1 and 2 exhibit striking resemblance.
In Zn-doped HTSC systems [4], 
the superfluid density $n_{s}/m^{*}$ at
$T\rightarrow 0$ decreases with increasing Zn concentration as shown in 
Fig. 3(a).  To explain this result, we proposed 
a ``swiss cheese model'' [4],
where each Zn suppresses superconductivity of the surrounding region 
characterized by the in-plane coherence $\xi_{ab}$ on the CuO$_{2}$ planes,
as illustrated in Fig. 3(d).  The solid lines represent the
expected superfluid density estimated from 
the ratio of superconducting versus non-superconducting regions.  
Without any fitting, this model gives a very good agreement with the
experimental data.  Recently this picture was confirmed directly by 
the scanning tunnelling microscope studies of Pan et al. [18].

\begin{figure}[t]
\epsfxsize=30pc 
\epsfbox{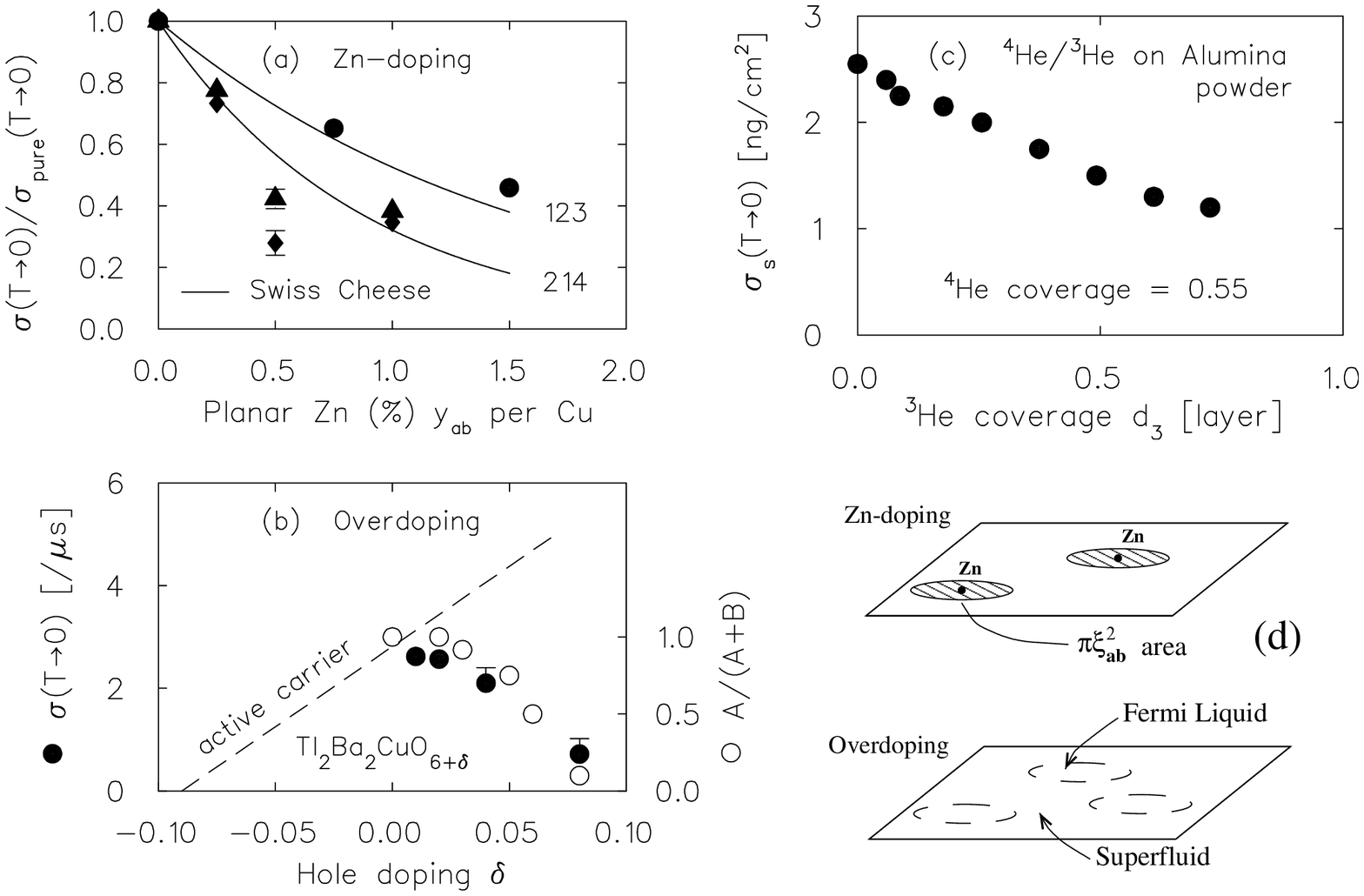} 
\caption{Depletion of the superfluid density due to perturbation:
(a) Zn-doping in YBa$_{2}$Cu$_{3}$O$_{6.63}$, 
La$_{1.85}$Sr$_{0.15}$CuO$_{4}$ and La$_{1.8}$Sr$_{0.2}$CuO$_{4}$ [4]; 
(b) Overdoping in Tl$_{2}$Ba$_{2}$CuO$_{6+\delta}$ [3]; and (c)
$^{3}$He mixing in $^{4}$He film adsorbed on fine alumina powder
[16].  
(d) illustrates the Swiss cheese model with Zn doping [4], and
proposed phase separation in overdoped HTSC. 
Open circles in (b) represents the relative weight of the 
gapped response (A) normalized to the sum of gapped (A) and 
ungapped (B) responses in the linear-$T$ term of the 
specific heat measurements by Loram {\it et al.\/} [20].}
\end{figure}

As shown in Fig. 1, $T_{c}$ of the Zn-doped cuprates follow the 
trajectory of hole-doped cuprates without Zn [4].  
This suggests that Zn
reduces $n_{s}/m^{*}$, and the volume average value of 
$n_{s}/m^{*}$ then determines $T_{c}$.  This situation looks quite   
analogous to $^{4}$He films adsorbed on porous media, where a part of He
forms a normal layer (i.e. a ``healing layer'') between the porous 
substrate and superfluid, while
$T_{c}$ is determined by the amount of the superfluid portion.

Recently, Kojima {\it et al.\/} [5] found that 
La$_{1.75}$Eu$_{0.1}$Sr$_{0.15}$CuO$_{4}$ (LESCO) undergoes magnetic order with 
static stripe freezing occurring in about half of the volume fraction
below $T_{N} \sim 10$ K.  This system becomes superconducting below 
$T_{c} \sim 30$ K.  The superfluid density could be determined 
even below $T_{N}$, thanks to the signal from the 
remaining non-magnetic volume.
The results show that $n_{s}/m^{*}$ is reduced to about a half of the 
value for La$_{1.85}$Sr$_{0.15}$CuO$_{4}$ without stripe formation.
This is consistent with a picture where hole carriers in the regions
with the frozen static spin stripes do not participate in the 
superfluid.  As shown in Fig. 1, $T_{c}$ again scales with 
the volume averaged value of $n_{s}/m^{*}$ in LESCO with static 
stripes.  Similar scaling with the trends of other 214
cuprates has been found in the $\mu$SR results of 
$T_{c}$ versus $n_{s}/m^{*}$ in single crystals of La$_{2}$CuO$_{4.12}$ and 
La$_{1.88}$Sr$_{0.12}$CuO$_{4}$ [6],
both of which having stripe spin freezing detected in a partial
volume fraction of muon sites.
The region with frozen stripes, though its size and origin are
yet to be clarified, looks analogous to the non-superconducting region
around Zn in Zn-doped cuprates.   
 
In overdoped Tl2201, $\mu$SR studies [3,19] revealed that $n_{s}/m^{*}$ 
{\it decreases\/} with increasing hole doping,
as shown in Fig. 3(b).  Since no signature of 
anomalous behavior has been found for $m^{*}$, this result 
suggests that $n_{s}$ becomes smaller than the normal state 
carrier concentration $n_{n}$.  Indeed a specific heat study in 
Tl2201 [20]
suggests co-existence of gapped (A) and un-gapped (B) responses, 
with the latter
portion increasing with increasing doping.  These results can be
explained if we assume a microscopic phase separation between 
superfluid and non-superconducting fermionic carriers [3,8,9,21], as
illustrated in Fig. 3(d).

A mixture of $^{4}$He and $^{3}$He provides a typical example of phase
separation.  With an increasing fermionic portion of $^{3}$He, $T_{c}$
decreases, maintaining an approximate proportionality to $p_{4}^{2/3}$
where $p_{4}$ denotes the volume fraction of $^{4}$He.  Thin films of 
$^{4}$He/$^{3}$He can be adsorbed on porous media, such as 
fine alumina powders, which constrain the phase separation to be 
microscopic.  In Fig. 3(c), we show the reduction of superfluid
density with increasing $^{3}$He fraction $p_{3}$ using
the results in ref. [16].  In this case, 
due to the 2-d configuration, $T_{c}$ decreases approximately as
$T_{c} \propto p_{4} = (1-p_{3})$.  Thus, both in 
$^{4}$He/$^{3}$He and in overdoped Tl2201, we see a suppression of the 
superfluid density and $T_{c}$ due to increasing fermionic fraction
in a microscopic phase separation. 

Based on these analogies, we propose a new phase diagram for 
HTSC systems in Fig. 4.  As stated in our previous publications 
[7-9,21],
we consider the ``pseudo-gap''
temperature $T^{*}$ to represent 
a signature of pair (boson) formation.  In this case,
$T^{*}$ reflects the magnitude of the attractive
interaction between fermionic carriers.  If this attractive 
interaction rapidly decreases with increasing hole doping
near the ``optimal $T_{c}$'' region, there is no robust 
superconductivity in the overdoped region.  However, the system can 
phase separate into regions with ``optimal hole density (OHD)'' 
(corresponding to ``optimal $T_{c}$'') which maintain
superconductivity and those with higher hole density (HHD) without 
superconductivity.  The charge imbalance will cost extra energy 
to phase separate while
bulk superconductivity could gain condensation energy.  The OHD
region would correspond to the $^{4}$He-rich superfluid while HHD region 
to the $^{3}$He-rich normal fluid in analogy to $^{4}$He/$^{3}$He.

\begin{figure}[t]
\begin{center}
\epsfxsize=20pc 
\epsfbox{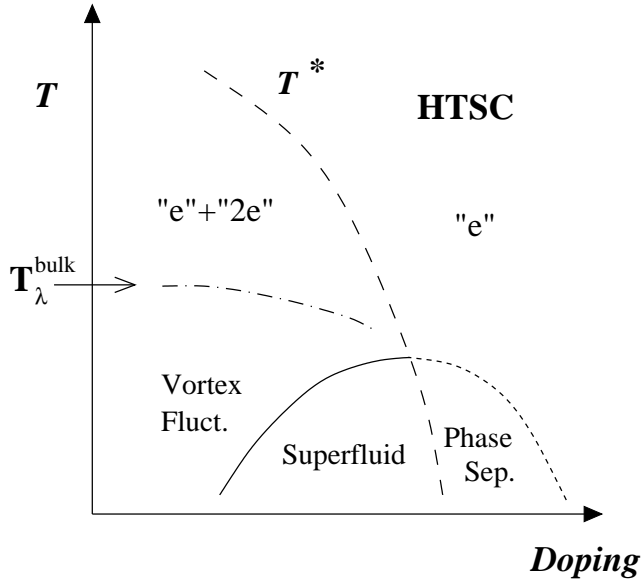} 
\caption{Proposed new phase diagram for HTSC systems.  
With decreasing temperature in the underdoped side, individual 
fermion carriers starts forming pairs below $T^{*}$,
and time-dependent superconductivity appears below $T_{\lambda}$.
In the overdoped side, there is no pairing interaction, and
superconductivity survives via phase separation.}
\end{center}
\end{figure}

Recently Loram, Tallon and co-workers [22] noticed 
a sharp reduction of the $T^{*}$ line
near the hole concentration of $x\sim 0.19$ per Cu.  They argued that this
phenomenon is incompatible with superconductivity observed at $x\geq 0.19$ in 
the overdoped region, if $T^{*}$ represents the superconducting
pairing interaction.  However, our picture with phase separation 
provides a way to reconcile the sharp reduction of the pairing 
energy scale $T^{*}$ with 
the survival of superconductivity in the overdoped region.
 
Signatures of fluctuating superconductivity have been 
found in the underdoped cuprates above $T_{c}$ in studies of 
high-frequency optical conductivity $\sigma_{ac}$ [23], 
the Nernst effect [24],
and the ``resonance'' inelastic scattering intensity in 
neutron scattering [25].  We notice that all these results show
onset of their effect below $T \sim 150$ K, which is substantially
lower than $T^{*}$ determined from the c-axis conductivity 
[26] and/or NMR
Knight shift [27].  The analogy to He films can provide a possible 
explanation to this feature.  

In the case of He, formation of bosons
(He atoms) from fermions occurs at very high temperatures.  
At a much lower temperature $T_{\lambda}^{bulk} = 2.2$ K 
(the bulk $\lambda$ transition point)
BE condensation occurs in a 3-d environment.  In a highly 2-d environment,
the bulk superfluidity occurs at lower temperature $T_{c}$.  
Time-dependent superfluidity via un-bound vortices occurs between
$T_{c}$ and $T_{\lambda}$ in the 2-d situation.  Similarly to this,
we expect the two step process, i.e., pair formation at a high 
temperature $T^{*}$ followed by signatures of fluctuating superconductivity
at much lower temperature below $T_{\lambda}$ ($T_{c} < T_{\lambda} < T^{*}$)
for underdoped cuprates, as illustrated in Fig. 4.  
We can ascribe the $\sigma_{ac}$, Nernst, and neutron results to 
the onset of time-dependent superconductivity below $T_{\lambda}$ while
the c-axis transport and Knight shift to the formation of singlet fermion
pairs below $T^{*}$.  Formation of a pair (boson)
does not immediately correspond to quantum condensation, which requires
a certain density/mass to achieve phase coherence of bosonic wave functions.

Despite all these analogous features, there exists an important 
difference between HTSC and He films.  Figure 5 shows a plot of
$T_{c}$ versus the  
2-d area superfluid density $n_{s2d}/m^{*}$ obtained for
cuprates by multiplying $n_{s}/m^{*}$ 
with the average distance $c_{int}$
between the CuO$_{2}$ planes.  
We notice that: (A) $T_{c}$ for the cuprates are 2-4 times reduced 
from $T_{KT}$ calculated for the strong-coupling limit;
(B) for a given $n_{s2d}/m^{*}$, $T_{c}$ is higher for systems with 
smaller $c_{int}$ (consistent with the results in inset for 
YBCO-PBCO multilayer films).  The feature (B) is incompatible with
the KT transition in pure 2-d systems where $T_{c}$ should not
depend on a 3-d coupling via $c_{int}$.  Previously, we 
pointed out that BE condensation in quasi 2-d systems would 
provide a better account for the observed dependence of 
$T_{c}$ on $c_{int}$
[9,21].

\begin{figure}[t]
\begin{center}
\epsfxsize=27pc 
\epsfbox{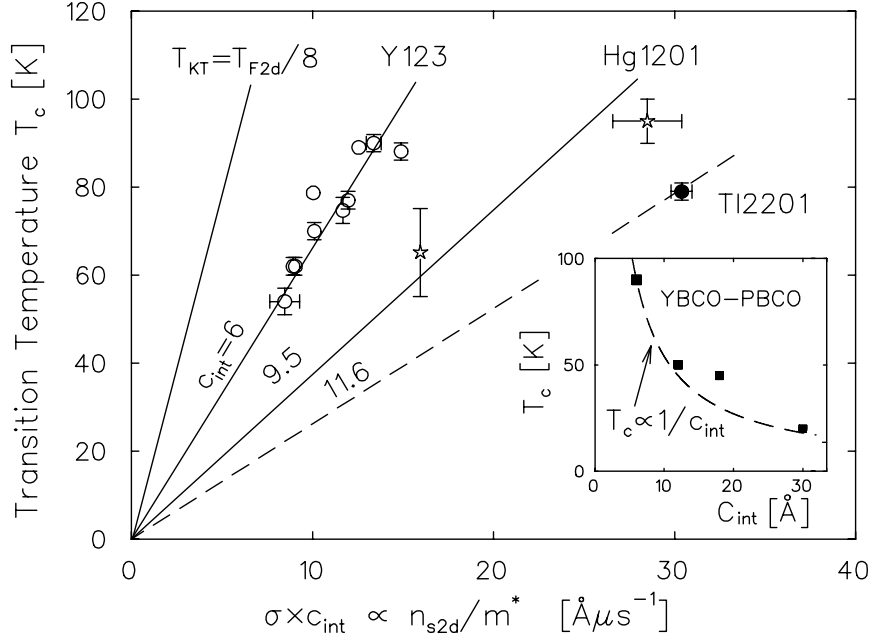} 
\caption
{Plot of $T_{c}$ versus $\sigma \times c_{int}
\propto n_{s2d}/m^{*} \propto T_{F2d}$ in underdoped 
and nearly optimally-doped HTSC [1,3,28] systems. 
$T_{KT}$ shows the strong coupling limit in pure 2-d.
Inset: $T_{c}$ vs. $c_{int}$ in multilayer YBCO-PrBCO films [29].} 
\end{center}
\end{figure}

Finally, we would like to point out that fluctuating
superconductivity can be expected not only for the KT
transition but also for BE-condensation in quasi 2-d systems.
Analogous, for example, to spin fluctuations in quasi 2-d
magnetic systems, correlations develop already at 
the temperature corresponding the transition temperature $T_{c3d}$
for a 3-d environment, while long-range order occurs 
at a much lower temperature $T_{c2d}$ due to dimensionality effects.
The correlated spin fluctuations at $T_{c2d} < T < T_{c3d}$
correspond to the fluctuating superfluidity in HTSC and $^{4}$He.
In BE condensation in the quasi 2-d situation, $T_{c}$ is 
determined at a point where thermal energy becomes
comparable to the interlayer interaction enhanced by the 
fluctuating in-plane superconducting correlations.  This
process is essentially similar to how $T_{c}$ for magnetic order is determined
in quasi 2-d spin systems.
Thus, the $\sigma_{ac}$, Nernst, and neutron results cannot
distinguish between a pure KT transition versus quasi 2-d
BE condensation.  This point requires further studies.

\section*{Acknowledgement}
The author is grateful to M. Randeria for helpful discussions.
This study is supported by NSF (DMR-98-02000) and US-Israeli Binational
Science Foundation.

\end{document}